\documentclass[%
prd,
reprint,nodate,
superscriptaddress,
 amsmath,amssymb,
 aps,
]{revtex4-1}

\pdfoutput=1

\usepackage[utf8]{inputenc} 

\usepackage{xcolor}
 
\usepackage{graphicx}
\usepackage{dcolumn}
\usepackage{bm}
\usepackage{microtype}
\usepackage{threeparttable} 
\usepackage{mathtools}
\usepackage[colorlinks=true]{hyperref}

\newcommand{\mysection}[1]{{\vspace{10 pt}\noindent \emph{{\textbf{#1}}--}}}

\newcommand{\rf}[1]{Eq.~(\ref{#1})}

\newcommand{\be}{\begin{eqnarray}}
\newcommand{\ee}{\end{eqnarray}}

\begin{document}

\author{Michal P.\ Heller} \email{michal.p.heller@aei.mpg.de}
\affiliation{Max Planck Institute for Gravitational Physics (Albert Einstein Institute),
14476 Potsdam-Golm, Germany}
\affiliation{National Centre for Nuclear Research, 02-093 Warsaw, Poland}
\affiliation{Department of Physics and Astronomy, Ghent University, 9000 Ghent, Belgium}

\author{Alexandre Serantes}
\email{alexandre.serantes@ub.edu}
\affiliation{National Centre for Nuclear Research, 02-093 Warsaw, Poland}
\affiliation{Departament de Física Quàntica i Astrofísica, Institut de Ciències del Cosmos (ICCUB), Facultat de Física, Universitat de Barcelona, Martí i Franquès 1, ES-08028, Barcelona, Spain}

\author{Micha\l\ Spali\'nski}
\email{michal.spalinski@ncbj.gov.pl}
\affiliation{National Centre for Nuclear Research, 02-093 Warsaw, Poland}
\affiliation{Physics Department, University of Bia{\l}ystok, 15-245 Bia\l ystok, Poland}

\author{Viktor Svensson}
\email{viktor.svensson@aei.mpg.de}
\affiliation{National Centre for Nuclear Research, 02-093 Warsaw, Poland}
\affiliation{Max Planck Institute for Gravitational Physics (Albert Einstein Institute),
14476 Potsdam-Golm, Germany}

\author{Benjamin Withers}
\email{b.s.withers@soton.ac.uk}
\affiliation{Mathematical Sciences and STAG Research Centre, University of Southampton, Highfield, Southampton SO17 1BJ, United Kingdom}

\title{Hydrodynamic Gradient Expansion Diverges beyond Bjorken Flow}

\begin{abstract}
The gradient expansion is the fundamental organising principle underlying relativistic hydrodynamics, yet understanding its convergence properties for general nonlinear flows has posed a major challenge. We introduce a simple method to address this question in a class of fluids modelled by Israel-Stewart--type relaxation equations. We apply it to (1+1)-dimensional flows and provide numerical evidence for factorially divergent gradient expansions. This generalises results previously only obtained for (0+1)-dimensional comoving flows, notably Bjorken flow. We also demonstrate that the only known nontrivial case of a convergent hydrodynamic gradient expansion at the nonlinear level relies on Bjorken flow symmetries and becomes factorially divergent as soon as these are relaxed. Finally, we show that factorial divergence can be removed using a momentum space cutoff, which generalises a result obtained earlier in the context of linear response.
\end{abstract}

\maketitle

\mysection{Introduction} Hydrodynamics plays a pivotal role in the description of nonequilibrium phenomena, with applications ranging from condensed matter systems~\cite{graphenefluid}
to scenarios in astrophysics~\cite{Shibata:2017jyf, Alford:2017rxf, Bemfica:2019cop} 
or nuclear physics~\cite{Heinz:2013th,Busza:2018rrf}. The reason is that hydrodynamics captures the infrared behavior of any medium endowed with conserved quantities. For a given set of conserved currents, the expression of hydrodynamic behaviour rests on the derivative expansion in the spirit of an effective field theory~\cite{Kovtun:2012rj,Hartnoll:2016apf,Florkowski:2017olj,Romatschke:2017ejr,Soloviev:2021lhs}. 
For a neutral relativistic fluid, the natural choice of dynamical variables are the energy density $\mathcal{E}(x)$ and the unit-normalized fluid velocity $U(x) = U^\mu(x) \partial_\mu$, with the conserved currents, $T^{\mu\nu}$, given
by the constitutive relation
\begin{equation}
\label{intro:T_decomposition}
T^{\mu\nu} = \mathcal{E} \, U^\mu U^\nu + P(\mathcal{E}) (g^{\mu\nu}+U^\mu U^\nu) + \Pi^{\mu\nu}. \end{equation}
Here, the first two terms describe ideal flow with $g$ being the Minkowski metric and $\Pi^{\mu \nu}$ captures dissipative effects organised as
\begin{equation}\label{intro:T_decomposition_3}
\Pi_{\mu\nu} = \sum_{n=1}^{\infty} \epsilon^n \Pi_{\mu\nu}^{(n)}[\mathcal{E},U],
\end{equation}
where $\Pi_{\mu\nu}^{(n)}$ contains $n$ spacetime derivatives of ${\cal E}, U$ and we have introduced $\epsilon$ as a formal derivative-counting parameter. The gradient expansion in \rf{intro:T_decomposition_3}  is defined up to redundancies associated with frame choice and current conservation $\nabla_\mu T^{\mu\nu} = 0$. 

Understanding the character of the expansion~\eqref{intro:T_decomposition_3} constitutes a fundamental open problem. Is it convergent, in such a way that subsequent truncations are progressively more accurate? If not, how does its divergent nature relate to the empirical 
success of low-order truncations?

Studies of comoving flows in Refs.~\cite{Heller:2013fn, Buchel:2016cbj,Casalderrey-Solana:2017zyh,Baggioli:2018bfa,Buchel:2018ttd,Aniceto:2018uik,Heller:2015dha,Basar:2015ava,Aniceto:2015mto,Denicol:2016bjh,Florkowski:2016zsi,Heller:2016rtz,Denicol:2017lxn,Heller:2018qvh,Blaizot:2019scw,Denicol:2019lio,Behtash:2017wqg,Denicol:2018pak,Behtash:2019qtk,Du:2021fok}, in which all fluid flow lines can be mapped to each other under symmetry transformations, rendering the problem effectively $(0{+}1)$-dimensional, have been instrumental in advancing our understanding of the hydrodynamic expansion~\eqref{intro:T_decomposition_3}. Among these, Bjorken flow~\cite{Bjorken:1982qr} in conformally invariant theories is the most thoroughly explored example due to its role in studies of quark-gluon plasma.
In these cases a particular strategy to solve the dynamical equations is an expansion in the Knudsen number, $1/w$ \cite{Heller:2011ju}. It is possible to compute a sufficient number of terms to assess that these expansions are factorially divergent. The expansion in $\epsilon$ defined in \rf{intro:T_decomposition_3} encapsulates the expansion in $1/w$, as we review in the Supplemental Material. Another well-studied example of a comoving flow is the Gubser flow which is reached by Weyl transformation from a ($0+1$)-flow on dS$_3\times \mathbb{R}$ \cite{Gubser:2010ze, Gubser:2010ui}. 

Outside the realm of comoving flows, the only generic result on~\eqref{intro:T_decomposition_3} was restricted to the linear response regime~\cite{Heller:2020uuy}. It showed that depending on how the momentum space support $k_\textrm{max}$ of $\mathcal{E}$ and~$U^\mu$ compares to an intrinsic scale of the underlying microscopic theory~$k_*$~\cite{Withers:2018srf}, the gradient-expanded constitutive relations could either be convergent ($k_\textrm{max}<k_*$), geometrically divergent ($k_*<k_\textrm{max}<\infty$) or factorially divergent ($k_\textrm{max}\to\infty$). 

In this Letter, we break free both from the symmetry constraints of comoving flows and the conveniences of linearisation to address, for the first time, the large-order behavior of the hydrodynamic gradient expansion beyond comoving flows at the fully nonlinear level. Specifically, 
we ask the following question: Given a generic nonequilibrium, nonlinear configuration of $\mathcal{E}$,~$U^\mu$ arising on-shell, what is nature of the expansion in $\epsilon$ \eqref{intro:T_decomposition_3} when evaluated on this solution? 
We answer this question by introducing a simple method that allows one to calculate~\rf{intro:T_decomposition_3} up to high order on a desktop computer. In this Letter we illustrate it in two examples. The first is the model put forward by Baier, Romatchske, Son, Starinets and Stephanov (BRSSS) in Ref.~\cite{Baier:2007ix}, while the second is the model originally introduced by Denicol and Noronha (DN) in Ref.\,\cite{Denicol:2019lio}. Both theories are representative examples within a broad class of models employing the Israel-Stewart approach to embed hydrodynamics in a framework compatible with relativistic causality~\cite{Israel1976Sep,Israel:1979wp}. Our method applies to any member of this class, such as~\cite{Denicol:2012cn}, and covers also equations with more than one derivative of $\Pi^{\mu\nu}$, such as~\cite{Jaiswal:2013vta,Noronha:2011fi,Heller:2014wfa}. Each member of this class generally gives rise to infinitely many transport coefficients in the gradient expansion~\eqref{intro:T_decomposition_3} that are specific to it.

\begin{figure*}
 \includegraphics[width=\textwidth]{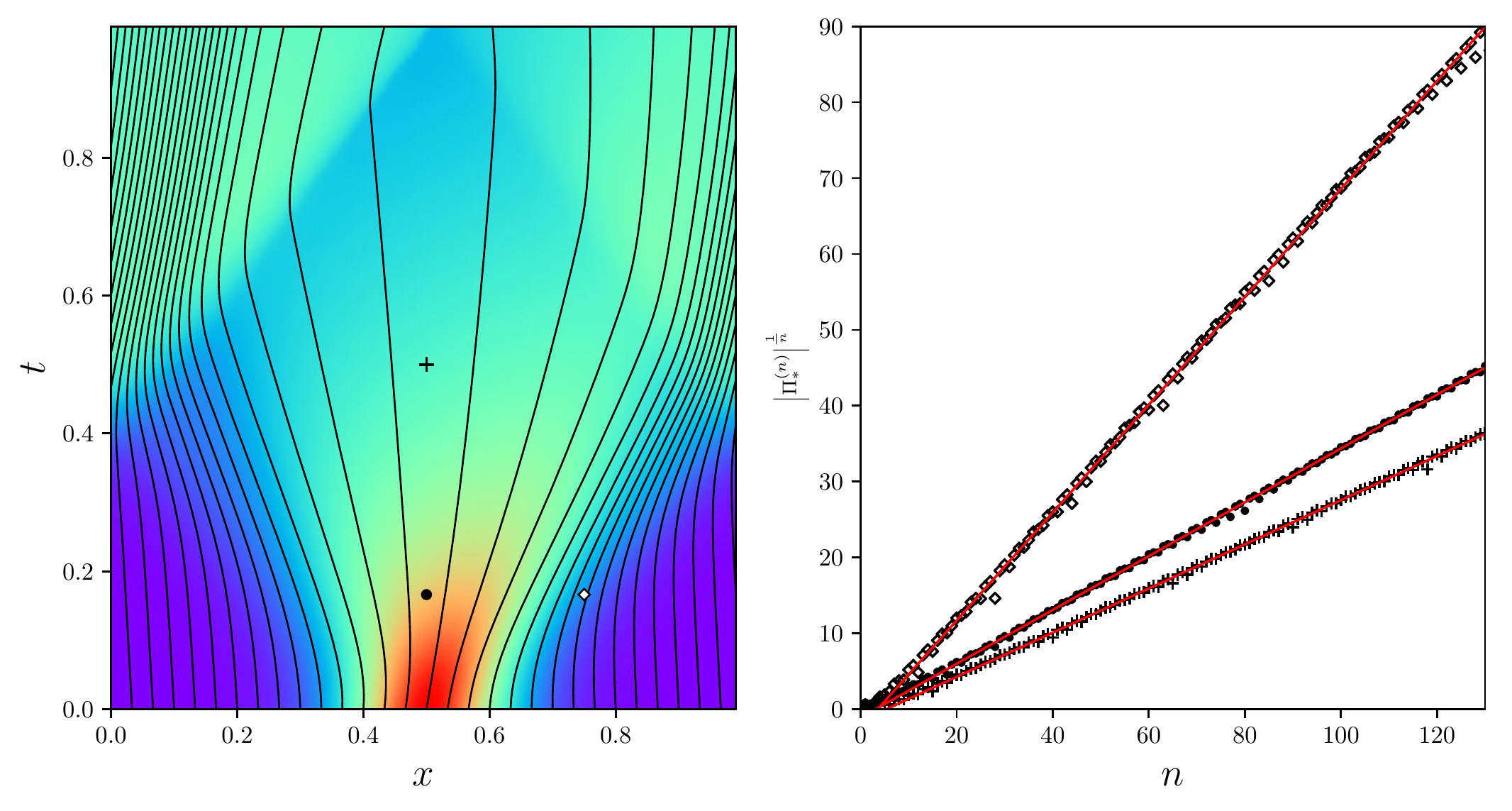}
 \caption{\textbf{Left panel: } Solution to an initial value problem in BRSSS. Colour shows the temperature profile~$T(t,x)$, and the black solid lines are flow lines of velocity~$U$. Here $t,x$ are obtained from standard Minkowski coordinates by periodically identifying $x\sim x+1$. Initial conditions were provided at $t=0$ corresponding to a strong Gaussian overdensity. The colour scale ranges from $\min{T} = 0.76$ (violet) to $\max{T} = 1.76$ (red). Also included are three marked points (disk, plus, open diamond). \textbf{Right panel:} Root plot for the hydrodynamic expansion of the constitutive relation \eqref{intro:T_decomposition_3} for BRSSS, evaluated using the recursion relation \eqref{BRSSS:recursion_relation} for the solution shown in the left panel, at each of the marked points with matching marker shapes. To guide the eye, in red we show straight lines fit to the range $n\in [20,130]$, which correspond to $\Pi_{\star}^{(n)}\sim \Gamma(n)$ at large $n$. Numerical convergence of our results is demonstrated in the Supplemental Material.}
 \label{heatmap}
\end{figure*}

\mysection{Results in the BRSSS model} In this Letter, we consider the restriction of the BRSSS model to conformal fluids in $d$ spacetime dimensions ($d = 4$ in our numerics). We work in the Landau frame and take $\Pi_{\mu\nu} U^\nu = 0$. The model is defined by promoting $\Pi_{\mu\nu}$ to a set of independent degrees of freedom subject to a relaxation equation,
\begin{equation}\label{BRSSS:dynamical_constitutive_relation}
\Pi^{\mu\nu} = -2 \eta \, {\cal D}^{\langle \mu}u^{\nu \rangle} - \tau_\Pi U^\alpha \mathcal{D_\alpha}\Pi^{\mu\nu} +\frac{\lambda_1}{\eta^2} \Pi^{\langle \mu}_\lambda \Pi^{\nu\rangle\lambda},
\end{equation}
where we neglected terms not relevant to the flow we consider. $\mathcal{D}_\mu$ is the Weyl-covariant derivative \cite{Loganayagam:2008is}, and the angle brackets instruct one to take the symmetric, transverse and traceless part of the tensor they act upon. The relaxation time $\tau_\Pi$, the shear viscosity $\eta$, and $\lambda_1$ are transport coefficients. Conformal invariance demands that these quantities depend on the local temperature $T$, defined by the relation $\mathcal{E} = \mathcal{E}_0 T^{d}$, as 
\begin{equation}
\eta = C_\eta \frac{4 \, {\cal E}}{3\, T}, \quad \tau_\Pi = \frac{C_{\tau_\Pi}}{T}, \quad \lambda_1 = C_{\lambda_{1}} \frac{\eta}{T}
\end{equation}
where $C_\eta$ and~$C_{\tau_\Pi} > 0 $. The equations of motion of BRSSS theory are given by  \eqref{BRSSS:dynamical_constitutive_relation} and the conservation equation $\nabla_\mu T^{\mu\nu} = 0$, where the energy-momentum tensor is specified in terms of $\mathcal{E}$, $U$ and $\Pi_{\mu\nu}$ as in~\eqref{intro:T_decomposition}. 

The fluid flows we focus on are characterised as follows. We separate the spatial coordinates into one longitudinal direction, $x$, and $d{-}2$ transverse directions, $x_1, \ldots, x_{d-2}$, demanding isotropy and translational invariance in the transverse hyperplane spanned by $x_i$. Hence, the nontrivial dynamics is confined to the longitudinal plane spanned by $t$ and $x$,  and our fluid flows are $(1{+}1)$-dimensional. We refer to these fluid flows as \textit{longitudinal}. At the linearized level such flows would correspond to sound wave propagation. See, e.g., Ref.~\cite{Florkowski:2016kjj} for a study of longitudinal flows in a quark-gluon plasma context.\footnote{Note that longitudinal flows are in general distinct from comoving flows such as Bjorken and Gubser flow. While Bjorken flow can be found as a special case of a longitudinal flow, Gubser flow cannot since it features transverse dynamics as dictated by symmetries.}

The most general fluid velocity 
for a longitudinal flow is parameterised by a single degree of freedom, $u$, as
\begin{equation}
U = U^\mu \partial_\mu = \cosh u\,\partial_t + \sinh u\,\partial_x. 
\end{equation}
Furthermore, any tensor which is symmetric, transverse to $U^\mu$ and traceless is described in terms of a single additional degree of freedom that we pick as
\begin{equation}\label{BRSSS:Pi_star_definition}
\Pi^{\mu\nu} = (2-d)\, \Pi_\star \Sigma^{\mu\nu}
\end{equation}
where $\Sigma^{\mu\nu} \equiv g^{\mu\nu} + U^\mu U^\nu - \frac{d{-}1}{d{-}2} P_T^{\mu\nu}$ and $\Pi_\star = \frac{1}{d{-}2} P_T^{\mu\nu} \Pi_{\mu\nu}$, with $P_T^{\mu\nu}$ being the projector in the transverse directions. 

We now consider the expansion \eqref{intro:T_decomposition_3} applied to the conformal BRSSS model for longitudinal flows. 
This is facilitated by a numerical algorithm which makes a computation of \eqref{intro:T_decomposition_3} to large orders tractable.
Since $\epsilon$ counts derivatives it can be introduced by taking \eqref{BRSSS:dynamical_constitutive_relation} and replacing $\nabla_\mu \to \epsilon \, \nabla_\mu$ together with positing a perturbative ansatz for $\Pi_\star$, as follows,
\be\label{BRSSS:gradient_expansion_epsilon}
{\cal D}_\alpha \to \epsilon\,{\cal D}_\alpha, \quad \Pi_\star \to \sum_{n=1}^{\infty} \Pi_\star^{(n)} \epsilon^n.
\ee
This leads to the following recursion relation
\begin{subequations}\label{BRSSS:recursion_relation}
\begin{align}
&\Pi_\star^{(1)} = -\frac{2}{d{-}2} \eta \, P_{T}^{\mu \nu} {\cal D}_{\langle \mu} U_{\nu\rangle}, \\
&\Pi_\star^{(n+1)} = -\tau_\Pi (U \cdot \partial) \Pi_\star^{(n)} - \frac{d(\partial \cdot U)}{d{-}1}  \tau_\Pi \Pi_\star^{(n)} \nonumber \\ &-(d{-}3)\frac{\lambda_1}{\eta^2} \sum_{m=1}^{n} \Pi_\star^{(m)}\Pi_\star^{(n+1-m)}, \quad n > 1. 
\end{align}
\end{subequations}
Here, $\mathcal{E}$ and $U$ are 
not expanded in $\epsilon$. Therefore to proceed to evaluate  \eqref{BRSSS:recursion_relation} we must first find  $\mathcal{E}$ and $U$ for a given choice of flow. These (as well as the exact $\Pi_\star$) can be obtained by numerically solving the BRSSS equations of motion as an initial value problem without invoking an $\epsilon$ expansion. Once $\mathcal{E}$ and $U$ are known, the recursion relation~\eqref{BRSSS:recursion_relation} can be efficiently evaluated numerically to high orders. Careful consideration of resolution and precision is required, as this procedure involves high numbers of successive derivatives of the background solution $\mathcal{E}$ and $U$. This is further discussed in the Supplemental Material,
where we show that our numerical results are convergent. Our approach applies to the whole class of theories which build on the Israel-Stewart approach.

Note that solving the recursion relation~\eqref{BRSSS:recursion_relation} \emph{analytically} is prohibitively expensive due to the fast growth in the number of terms contributing at each order. In particular, for the case $\lambda_{1} = 0$ we observed an exponential growth of the number of individual contributions at each order. Our method circumvents this difficulty.

We have applied our approach in the BRSSS model across a wide variety of initial conditions, transport coefficients and spacetime points. We find factorial growth in all cases considered. We illustrate this in Fig.~\ref{heatmap} with one representative example in which we consider a strong Gaussian overdensity for our initial conditions and adopt a periodic compactification of the spatial direction. The left panel shows $T$ and flow lines of $U$ and highlights three spacetime point samples. The right panel root plot demonstrates factorial growth at these sampled points. Further details are provided in the Supplemental Material. 

\begin{figure}[h]
\begin{center}
\includegraphics[width=\columnwidth]{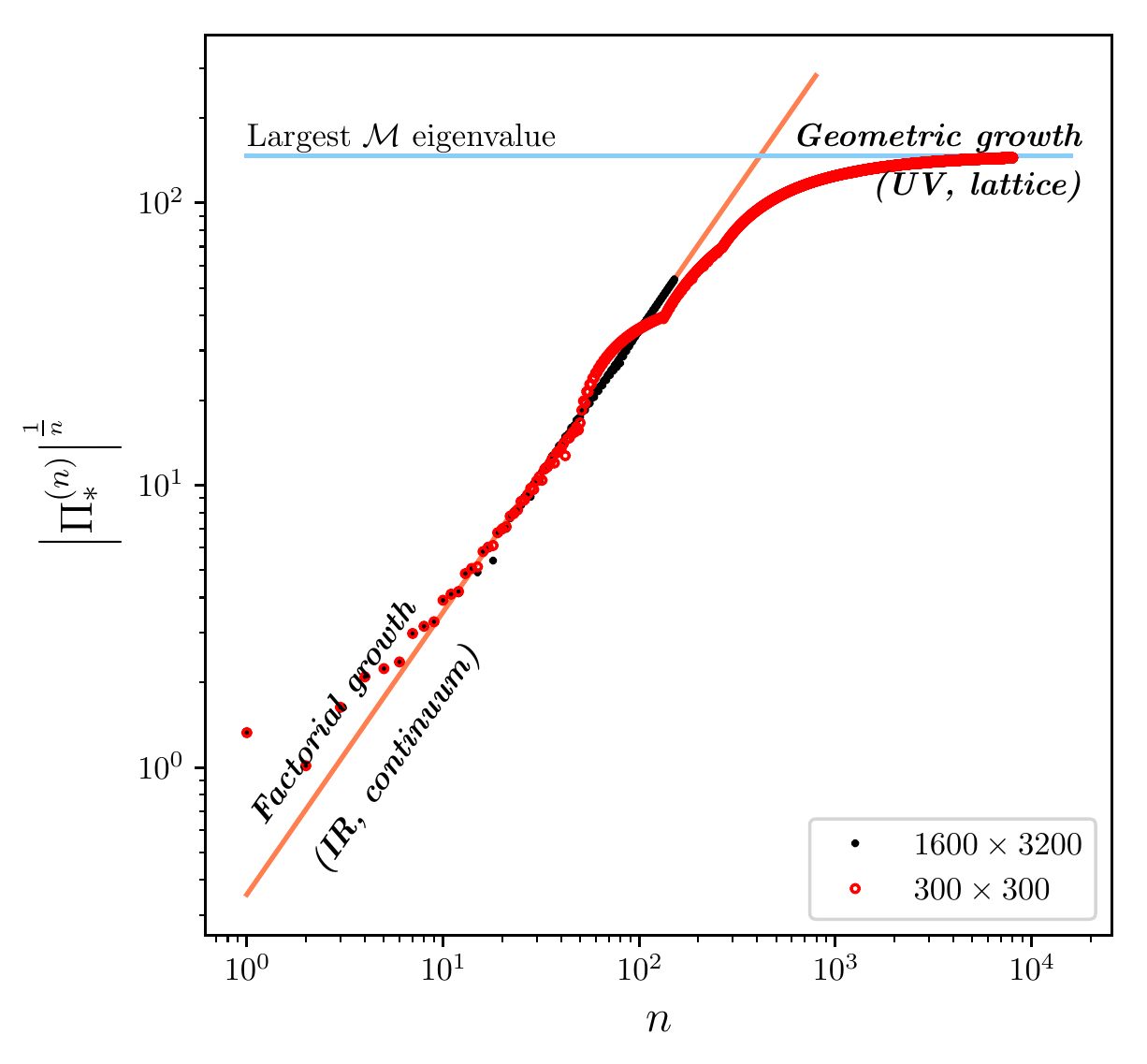}
\caption{Working on a lattice gives a window of factorial growth--where it successfully approximates the continuum--before yielding to a geometrically growing hydrodynamic expansion regulated by the lattice. The asymptotic value reached is governed by the lattice spacing as discussed in the text. Black disks are those from Fig. \ref{heatmap} corresponding to the disk spacetime marker point. Red circles are the same simulation and spacetime point, but on a coarse numerical grid and evaluated to hydrodynamic order $n=8000$. This plot serves also as an indication of a convergence of our approach.}
\label{fig:cutoff}
\end{center}
\end{figure}

\mysection{Momentum cutoff} In previous work \cite{Heller:2020uuy} we showed that a momentum-space cutoff gives at most a geometrically divergent hydrodynamic expansion for linear deviations from equilibrium. This result naturally extends to strongly nonlinear scenarios, as we now demonstrate. So far in this Letter we have used a numerical grid simply as a tool to approximate the continuum, but we now push beyond the continuum picture and re-evaluate the grid in a new role as a physical lattice which naturally enforces a momentum-space cutoff. In the BRSSS model, when $\lambda_1 = 0$ the recursion relation \eqref{BRSSS:recursion_relation} can be written as
\be
\Pi_\star^{(n+1)} = {\cal M}\, \Pi_\star^{(n)} \qquad n>1\label{M_def}
\ee
where ${\cal M} = -\tau_\Pi (U \cdot \partial) - \frac{d(\partial \cdot U)}{d{-}1}$ is a differential operator independent of $n$, depending only on the background solution ${\cal E}, U$. For a grid of dimensions $N_x\times N_t$, each $\Pi_\star^{(j)}$ can be written as a $N_xN_t$-sized vector, and ${\cal M}$ accordingly as a $N_xN_t \times N_xN_t$ square matrix.
Thus, on a lattice the expansion is ultimately only geometrically growing at a rate set by the largest eigenvalue of ${\cal M}$, which scales with the inverse lattice spacing. In Fig. \ref{fig:cutoff} this is demonstrated by utilising a deliberately low resolution lattice to allow for evaluating the hydrodynamic expansion to order $n=8000$. It shows the transition from factorial growth where the continuum approximation holds, to the geometrically divergent asymptotic behaviour governed by the aforementioned eigenvalue. We have also verified numerically that this result holds at $\lambda_1 \neq 0$, with the definition of ${\cal M}$ as given.

\mysection{Resolving the DN model tension} Bjorken flow is a boost-invariant longitudinal flow such that the dynamics depends on the proper time $\tau = \sqrt{t^2 - x^2}$ only. In Ref.~\cite{Denicol:2019lio} the authors analysed a Knudsen number expansion for an ultrarelativistic gas of hard spheres undergoing Bjorken flow. 
While in all the other models such expansions have been observed to be factorially divergent,
in Ref.~\cite{Denicol:2019lio} the 
terms grow only geometrically, with convergence ensuing for a Knudsen number smaller than a critical value. Our objective is to re-analyse this physical scenario using the expansion in $\epsilon$. For Bjorken flow we find analogous results, namely geometric growth; however, our method allows us to explore what happens when these symmetries are relaxed.

We work in $d=4$. As in the conformal BRSSS model, the energy-momentum tensor in the DN model is traceless and decomposed as in Eq.\,\eqref{intro:T_decomposition}, with $\Pi_{\mu\nu}$ still obeying Eq.\,\eqref{BRSSS:dynamical_constitutive_relation} with $\lambda_1 = 0$. Hence, the recursion relations giving $\Pi_\star^{(n)}$ still take the form \eqref{BRSSS:recursion_relation}, again with zero $\lambda_1$.  The differences start with the inclusion of a conserved current $J^\mu = \rho\, U^\mu$, where $\rho$ is the particle density. Furthermore, $\tau_\Pi$ and $\eta$ are not fixed purely in terms of the local temperature $T$, but rather obey 
\begin{equation}\label{DN:transport_coefficients}
\mathcal{E} = 3 \rho\,  T, \quad \eta =  \frac{a}{\sigma_T} T, \quad \tau_\Pi = \frac{a \, b}{4 \sigma_T}\frac{1}{\rho}, 
\end{equation}
where $\sigma_T$ is the total cross section and $a,\,b$ are positive dimensionless constants.   

For Bjorken flow, the conservation of the particle current $J^\mu$ entails that the particle density $\rho$ decouples from the energy-momentum tensor. One has that 
\begin{equation}
\rho(\tau) = \frac{\rho_0 \, \tau_0}{\tau},     
\end{equation}
where $\rho_0 = \rho(\tau_{0})$ is the initial particle density. Hence, 
\begin{equation}\label{DN:Bjorken_tau_Pi}
\tau_\Pi = \frac{1}{4} a \, b\, \textrm{Kn}\,\tau,    
\end{equation}
where $\textrm{Kn} = 1/(\rho_0 \tau_0 \sigma_T)$ is the Knudsen number. In the DN model for Bjorken flow it is time independent. 

To assess the large-order behavior of the expansion in $\epsilon$,  \rf{intro:T_decomposition_3}, we first note that one can find a closed-form expression for $\Pi_\star^{(n)}$,
\begin{align}\label{DN:Bjorken_recursion-relation_solution}
\Pi_\star^{(n)}(\tau) = &\frac{2}{3} a\,\textrm{Kn}\,\rho_0 \left( \frac{\tau_0}{\tau}\right)^\frac{4}{3} \left(- \frac{a b \textrm{Kn}}{4}\right)^{n-1} \times \nonumber \\
&\times (\tau \partial_\tau)^{n-1}\left(\left(\frac{\tau}{\tau_0}\right)^\frac{1}{3} T(\tau)\right),   
\end{align}
a fact that relies crucially on Eq.\,\eqref{DN:Bjorken_tau_Pi}. Second, we recall that $T$ can also be determined exactly \cite{Denicol:2019lio}
\begin{equation}\label{DN:Bjorken_closed-form_solution}
T(\tau) = T_{0,+}\left(\frac{\tau_0}{\tau}\right)^{\alpha_+} + T_{0,-}\left(\frac{\tau_0}{\tau}\right)^{\alpha_-}, 
\end{equation}
where $T_{0,\pm}$ depend on initial conditions and $\alpha_\pm$ on $a, b,$ and $\textrm{Kn}$. Together, Eqs.\,\eqref{DN:Bjorken_recursion-relation_solution} and \eqref{DN:Bjorken_closed-form_solution} entail that $\Pi_\star^{(n)}$ cannot grow factorially with $n$ at fixed $\tau$, since the repeated action of the differential operator $\tau \partial_\tau$ on terms of the form $(\tau/\tau_0)^{\frac{1}{3}-\alpha_{\pm}}$ only grows geometrically. Furthermore,  Eqs.\,\eqref{DN:Bjorken_recursion-relation_solution} and \eqref{DN:Bjorken_closed-form_solution} also imply that $\Pi_\star^{(1)}$ (and therefore all $\Pi_\star^{(n>1)}$) is a linear combination of eigenfunctions of the differential operator $\mathcal{M}$ defined as in Eq.\,\eqref{M_def}, providing another perspective on why the gradient expansion grows geometrically in this case. We refer the reader to the Supplemental Material for further details.

\begin{figure}[h]
\begin{center}
\includegraphics[width=\columnwidth]{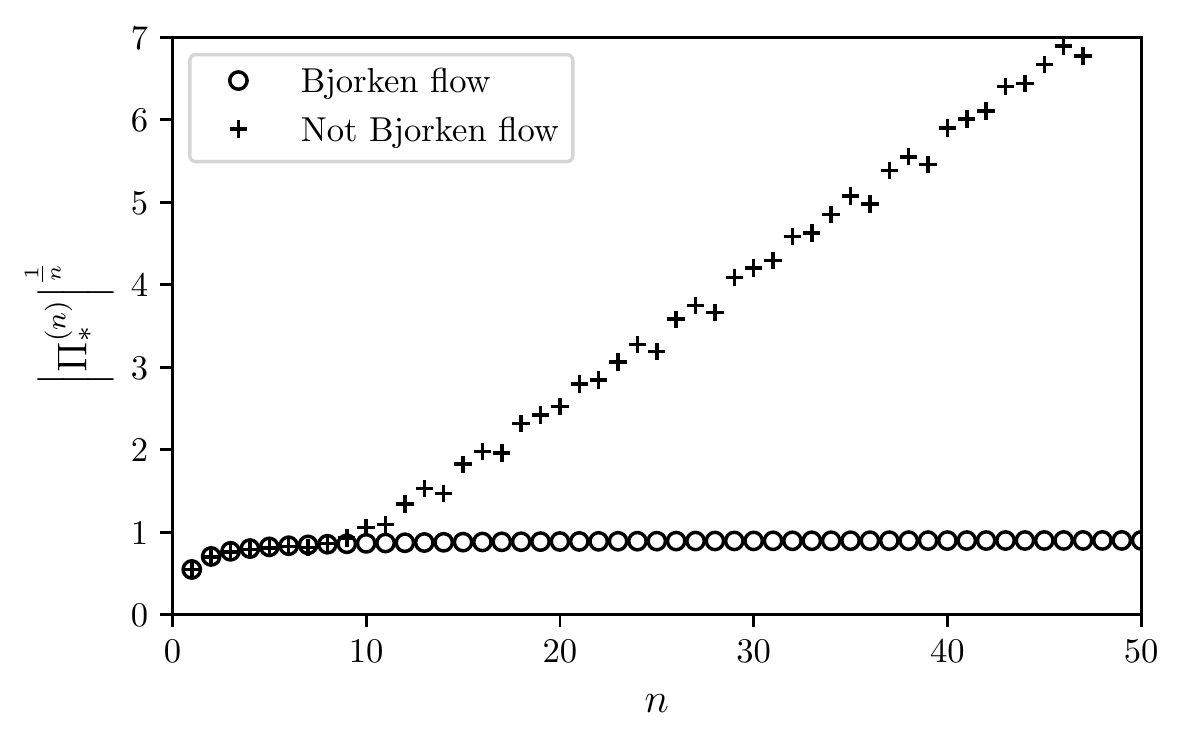}
\caption{Deviations from Bjorken symmetries restores factorial growth in the DN model. Open circles represent $|\Pi_\star^{(n)}|^\frac{1}{n}$ for Bjorken flow in the DN model evaluated at $\tau = 1$, where we have chosen initial conditions given by $\rho(\tau_0)=4$, $T(\tau_0) = 2$ and $\Pi_\star(\tau_0)/(\rho(\tau_0) T(\tau_0)) = 0.436461$ at $\tau_0 = 0.1$. Crosses represent $|\Pi_\star^{(n)}|^\frac{1}{n}$ at $\tau=1$ and zero rapidity for a longitudinal flow in the DN model which obeys the same initial conditions as Bjorken flow, but with a rapidity-even Gaussian overdensity in $\rho$ of amplitude $0.1$ and unit width. We have set $a=0.5$, $b=10$ and $\sigma_T=1$. The Bjorken flow coefficients have been obtained from the analytic solution, while the non-Bjorken flow ones from a simulation discussed in the Supplemental Material.}
\label{fig:DN_divergence}
\end{center}
\end{figure}

The analysis above relies crucially on the symmetry restrictions of Bjorken flow. Empirically, when relaxing these symmetry restrictions in all cases studied we find that the large-order geometric growth is destroyed and the factorial divergence is restored. We illustrate this in Fig.\,\ref{fig:DN_divergence} for a longitudinal flow corresponding to a small perturbation of Bjorken flow.  

\mysection{Summary and outlook} Understanding the behaviour of the  hydrodynamic gradient expansion at large orders is a challenging question in the  foundations of relativistic hydrodynamics. We have proposed a method to compute  such series in a large class of models. Applying this to nonlinear longitudinal flows reveals factorially divergent series  which we have illustrated with a number of examples. This shows that previously observed instances of factorial growth were not reliant on Bjorken symmetries.

It is natural to ask what are the generic conditions that lead to factorial growth. We have here established that the ability of a system to support arbitrarily large momentum is important; ways around this include working on a lattice, and appropriate restriction of initial data in the linearised case \cite{Heller:2020uuy}, both of which naturally lead to geometric growth. Second, as our analysis of the DN model shows, imposing special symmetries such as boost invariance can also lead to geometric rather than factorial growth. It would be interesting to explore this question further including other natural momentum cutoffs such as microscopic physics and turbulent cascades. For models with recursion relations of the form \eqref{M_def}, it may be possible to engineer further examples where the underlying equations of motion give rise to $\Pi^{(n)}_\star$ which are eigenfunctions of ${\cal M}$ such that the hydrodynamic gradient expansion grows geometrically. This could form the basis of a rigorous mathematical formulation for investigating the genericity of factorial growth.

The picture that is emerging from this work and results in linear response \cite{Heller:2020uuy} is that the origin of factorial growth at large $n$ in the hydrodynamic gradient expansion is the successive action of $n$ derivatives on the hydrodynamic variables ${\cal E}$,~$U$. This is intimately connected with support of a solution in momentum space. It suggests that having a factorial growth in the \emph{number} of transport coefficients at each order is not necessary.

The factorial divergence of asymptotic series is not an impediment to their practical utility: such series typically provide excellent approximations as long as one does not exceed the so-called order of optimal truncation. Our work makes such investigations possible for a much wider set of flows than previously tractable.

\mysection{Acknowledgements} The Gravity, Quantum Fields and Information group at AEI is supported by the Alexander von Humboldt Foundation and the Federal Ministry for Education and Research through the Sofja Kovalevskaja Award. A.\,S. was supported by the Polish National Science Centre grant 2018/29/B/ST2/02457 and by grant CEX2019-000918-M funded by MCIN/AEI/10.13039/501100011033. M.\,S. is supported by the National Science Centre, Poland, under grants 2018/29/B/ST2/02457 and 2021/41/B/ST2/02909. For the purpose of Open Access, the author has applied a CC-BY public copyright licence to any Author Accepted Manuscript (AAM) version arising from this submission. B.\,W. is supported by a Royal Society University Research Fellowship.

\bibliography{literature}
\bibliographystyle{bibstyl}

\onecolumngrid
\clearpage
\setcounter{page}{0}
\begin{center}
\textbf{\large Supplemental Material}
\vspace{1em}
\end{center}
\twocolumngrid
\appendix

\section{The expansions in $\epsilon$ and $1/w$ in Bjorken flow}\label{app:1overw}

In the studies of conformal Bjorken flow in the past decade, an important role was played by the following dimensionless clock variable~\cite{Heller:2011ju}
\begin{equation}
w \equiv \tau \, T(\tau),
\end{equation}
where $\tau$ is the proper time and $T(\tau)$ is the local temperature as defined in the main text. The purpose of this Appendix is to review the known relation between the expansion in the Knudsen number in conformal Bjorken flow -- the $1/w$-expansion mentioned in the Introduction -- and the gradient expansion of hydrodynamic constitutive relations as given by Eq.~\eqref{intro:T_decomposition_3} in the main text. 

While this relation is general, i.e. it applies to any conformally invariant model, we will present it in the context of the BRSSS model captured by Eq.~\eqref{BRSSS:dynamical_constitutive_relation} in the main text in four spacetime dimensions. For Bjorken flow Eq.~\eqref{BRSSS:dynamical_constitutive_relation} reduces to
\begin{align}
&\Pi_\star'(\tau) + \left(\frac{4}{3\tau} + \frac{T(\tau)}{C_{\tau_\Pi}}\right)\Pi_\star(\tau) - \frac{3 C_{\lambda_1}}{4 C_\eta C_{\tau_\Pi}\mathcal{E}_0 T(\tau)^3}\Pi_\star(\tau)^2 \nonumber \\
&- \frac{8 C_\eta \mathcal{E}_0 T(\tau)^4}{9 C_{\tau_\Pi} \tau} = 0. 
\end{align}
whereas the conservation equation always (in all theories) reads
\begin{equation}
T'(\tau) + \frac{T(\tau)}{3\tau} - \frac{\Pi_\star(\tau)}{2 \mathcal{E}_0 \tau T(\tau)^3}=0. 
\end{equation}
Note that it is convenient to replace $\Pi_{\star}$ by a dimensionless quantity by taking its ratio with the energy density~${\cal E}$. This is simply related to the pressure anisotropy $\cal A$ measuring deviations from local thermal equilibrium in the following way
\begin{equation}
{\cal A} = \frac{{\cal P}_T-{\cal P}_L}{{\cal E}/3} = 9 \frac{\Pi_{\star}}{\cal E}. 
\end{equation}
The $1/w$-expansion has been primarily discussed for the this quantity and up to second order it takes the form
\begin{equation}
\label{eq.Ain1/w}
{\cal A} = \frac{8 C_\eta}{w} + \frac{16 C_\eta (C_{\tau_\Pi}-C_{\lambda_1})}{3w^2}+ O\left(\frac{1}{w^3}\right).  
\end{equation}
Using the expansion in $\epsilon$ of $\Pi_{\star}$ at low orders we obtain the following expression for $\cal A$
\begin{align}
{\cal A} = \frac{8C_\eta}{\tau T}\epsilon {-}\frac{8 C_\eta \left((C_{\tau_\Pi} {+}2 C_{\lambda_1}) T{+}9 C_{\tau_\Pi} \tau T' \right)}{3 \tau^2 T^3}\epsilon^2{+} O(\epsilon^3),
\end{align}
where $T = T(\tau)$. One can now use the conservation equation to replace derivatives of $T(\tau)$ at each order in terms of powers of~$w$, which directly leads to~\eqref{eq.Ain1/w}. This shows  that knowing $\Pi_{\star}$ in the expansion in $\epsilon$ to order $n$ allows one to generate the expansion of $\cal A$ in $1/w$ up to and including terms $O(1/w^n)$. Note however, that the $n$-th order of the expansion in $1/w$ contains in principle contributions from the orders 2 to $n$ of the expansion in~$\epsilon$. 
A key difference between the two expansions is that the functional expansion of $\Pi_\star$ in derivatives of ${\cal E}, U$ counted by~$\epsilon$ as in  \rf{intro:T_decomposition_3} does not explicitly require invoking conservation laws (beyond the stated redundancies), whereas the $1/w$-expansion necessarily utilizes the conservation equation. Finally, note again that the discussion above applies to conformal Bjorken flow, while in Appendix \ref{app:DN} we discuss a particular non-conformal version of Bjorken flow. 

\section{Further details on numerics and convergence \label{app:num}}
For the BRSSS simulations shown in Fig.\,\ref{heatmap} we use initial data $u = \frac{1}{4} f$,~$T = 1 + f$,~$\Pi_\ast = 0$ where
\be
f(x) = e^{-\frac{\cos(\pi x)^2}{2\pi^2 \gamma^2}} -e^{-\frac{1}{4\pi^2 \gamma^2}}I_0\left(\frac{1}{4\pi^2 \gamma^2}\right)
\ee
on a unit spatial circle ($x\sim x+1$). $f$ was chosen to locally reproduce a Gaussian of width $\gamma$ near $x=1/2$, respect the spatial periodicity, and the additive constant chosen so that $f$ has no homogeneous component in a cosine expansion. We used 
${\cal E} = T^4$,~$C_{\tau_\Pi} = 1/4$,~$C_{\eta} = 1/(4\pi)$,~$\lambda_1 = 0$,~$\gamma^2 = 1/60$. The initial value problem is solved using RK4 on a uniform spatial grid with $N_x = 1600$ points for a total of $N_t=3200$ timesteps to reach the arbitrary choice of end time $t=1$, working with $300$ digits of precision. The differential operator for $\partial_x$ used periodic fourth-order finite differences, and second-order finite differences for $\partial_t$ one-sided at $t=0$ and $t=1$. 

\begin{figure}
\begin{center}
\includegraphics[width=\columnwidth]{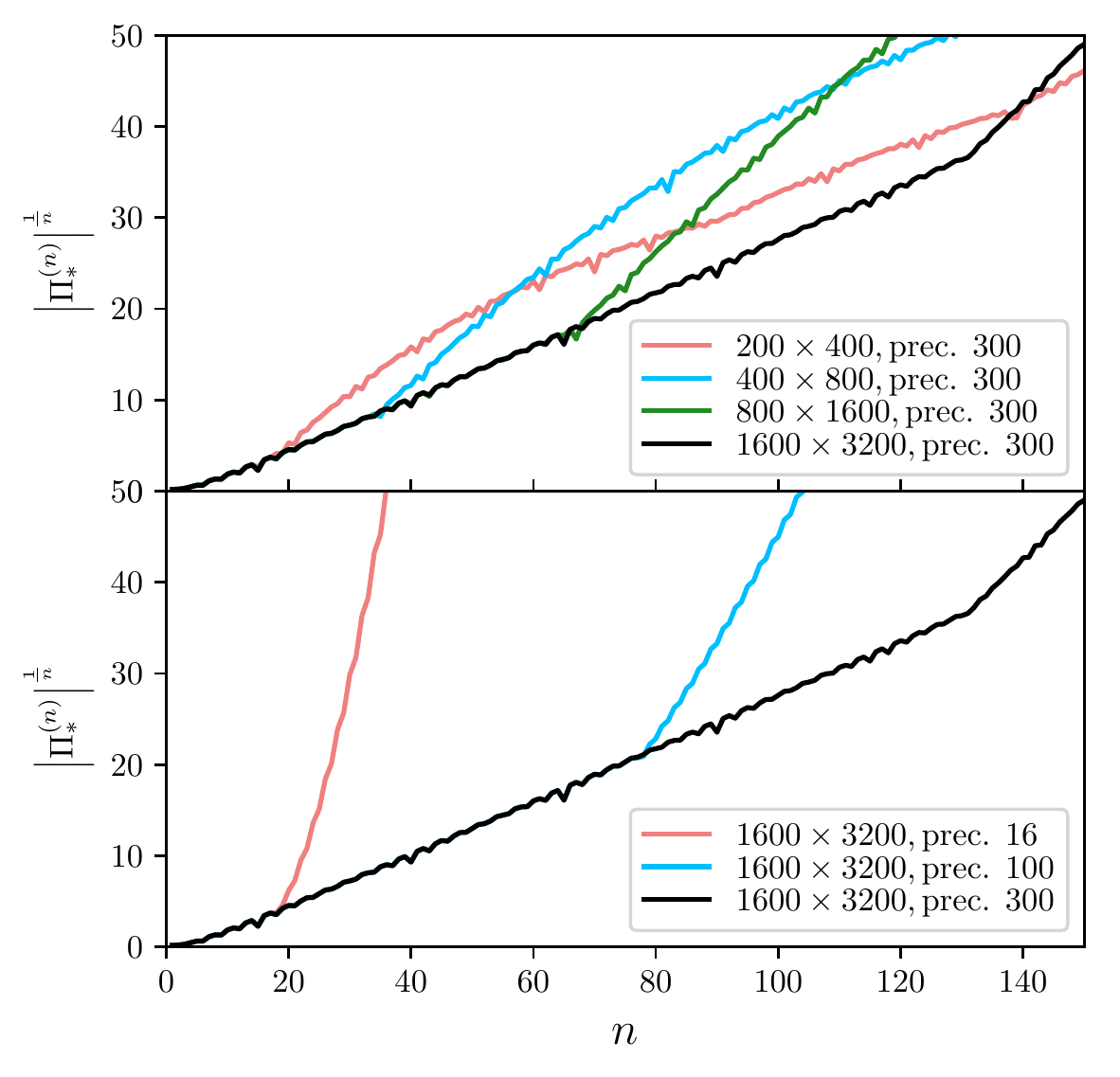}
\caption{Both precision and numerical resolution are important factors in extracting the correct growth of the hydrodynamic expansion using our technique. Here we show convergence of $|\Pi_\star^{(n)}|^{\frac{1}{n}}$ for the BRSSS simulations of Fig.\,\ref{heatmap} at the point marked with `$+$'. \textbf{Upper panel:} Convergence to the continuum with increasing resolution at fixed precision. \textbf{Lower panel:} Convergence with increasing precision at fixed resolution.}
\label{fig:convergence}
\end{center}
\end{figure}

Due to the large numbers of successive applications of derivative operators used in evaluating $\Pi_\star^{(n)}$ we have placed particular importance on testing convergence of our results, both in the resolution of the numerical grid $N_x,N_t$ and in the number of digits of precision used. Tests confirming convergence of $|\Pi_\star^{(n)}|^{\frac{1}{n}}$ are shown in Fig.\,\ref{fig:convergence}. For a given resolution and precision, when the window of factorial growth ends it can be extended by increasing one, the other, or both. 

The results in the DN model had been obtained working in a curvilinear coordinate system in which the Minkowski metric reads 
\begin{equation}\label{numerics:ds2_curvilinear_coordinates}
ds^2 = - e^{2\alpha(\tau,x)}d\tau^2+e^{2\beta(\tau,x)}dx^2 + d\vec{x}_\perp^2,     
\end{equation}
and the fluid velocity is $U = e^{-\alpha}\partial_\tau$, in such a way that a line of constant $x, \vec{x}_\perp$ corresponds to a flow line. Besides the conservation equations and the relaxation equation obeyed by $\Pi_\star$, the evolution equations include one extra equation enforcing that the metric \eqref{numerics:ds2_curvilinear_coordinates} is flat. When performing numerical simulations, we worked with a compactified spatial coordinate $\zeta \in [-1,1]$ instead of $x$. Both coordinates are related as $x = \gamma^{-1}\tanh^{-1}(\zeta)$, with $\gamma \in \mathbb{R}^+$. Spatial derivatives were discretized with fourth-order centered finite-difference stencils, while the time evolution was performed with an explicit RK4 method. For the numerical results displayed in Fig.\,\ref{fig:DN_divergence}, we employed a spatial grid of spacing $d\zeta = 1/1800$, a time step $d\tau =0.5 d\zeta$, and worked with precision $300$. The time derivatives appearing in the recursion relation were computed with fourth-order finite-difference stencils.

\section{Further perspectives on geometric growth in the DN model}\label{app:DN}

We start by noting that the DN model, as originally defined in Ref.\,\cite{Denicol:2019lio}, features an additional term in the relaxation equation obeyed by $\Pi_{\mu\nu}$. This term is linear in $\Pi_{\mu\nu}$ and, in this work, we have set to zero the transport coefficient associated to it since this restriction does not modify our main conclusions regarding Bjorken flow. Hence, in the formulation of the DN model we have considered, the gradient expansion \eqref{intro:T_decomposition_3} for Bjorken flow is given by the following recursion relations, 
\begin{subequations}\label{DN_rec_rel_tau}
\begin{equation}\label{DN_rec_rel_tau_1}
\Pi_\star^{(1)}{=}\frac{2a}{3\sigma_T} \frac{T}{\tau}, 
\end{equation}
\begin{equation}\label{DN_rec_rel_tau_2}
\Pi_\star^{(n+1)}{=}{-}\frac{a b \textrm{Kn}}{3}\Pi_\star^{(n)}{-}\frac{a b \textrm{Kn}}{4} \tau\partial_\tau \Pi_\star^{(n)}.    
\end{equation}
\end{subequations}
The closed-form expression \eqref{DN:Bjorken_recursion-relation_solution} can be shown to be a solution of the recursion relations \eqref{DN_rec_rel_tau} by induction. The parameters appearing in Eq.\,\eqref{DN:Bjorken_closed-form_solution} are 
\begin{subequations}\label{DN_T_exact_parameters}
\begin{equation}
T_{0,+} = \frac{c_0 \mathcal{E}_0}{3 n_0 (1+ c_0)} = c_0 T_{0,-},     
\end{equation}
\begin{equation}\label{DN_alpha_pm}
\alpha_\pm = \frac{1}{3} + \frac{2}{a b \textrm{Kn}} \pm \frac{2}{3}\sqrt{\frac{4}{b} + \frac{9}{a^2 b^2 \textrm{Kn}^2}},     
\end{equation}
\end{subequations}
where $\mathcal{E}_0$ is the energy density at $\tau=\tau_0$. Combining Eq.\,\eqref{DN:Bjorken_recursion-relation_solution} with Eq.\,\eqref{DN:Bjorken_closed-form_solution}, one finds the final form of the gradient expansion coefficients,
\begin{subequations}\label{DN_grad_exp_Bjorken}
\begin{equation}\label{DN_grad_exp_Bjorken_1}
\Pi_\star^{(n)} = \sum_{i=+,-} \Pi_{\star,i}^{(n)} = \sum_{i=+,-} A_i \left(\frac{\tau_0}{\tau}\right)^{4\frac{a b \textrm{Kn} + 3 \gamma_i}{3 a b \textrm{Kn}}}\gamma_i^n,    
\end{equation}
\begin{equation}
A_\pm = - \frac{8 T_{0,\pm}}{b \tau_0 \sigma_T \textrm{Kn} (1- 3\alpha_\pm)}, \end{equation}
\begin{equation}\label{DN_gamma_pm}
\gamma_\pm = \frac{1}{2} \pm\frac{1}{6}\sqrt{9+ 4 a^2 b \textrm{Kn}^2}.     
\end{equation}
\end{subequations}
According to Eq.\,\eqref{DN_grad_exp_Bjorken}, $\Pi_\star^{(n)}$ is given by the sum of two contributions which only grow geometrically at rates $\gamma_\pm$. No large-order factorially growing behavior is present. 

For illustrative purposes, let us focus on the $c_0=0$ solution. This choice of initial conditions corresponds to the attractor solution for the normalized pressure anisotropy, which played a prominent role in the original analysis of Ref.\,\cite{Denicol:2019lio}. In this case, the terms in Eq.\,\eqref{DN_grad_exp_Bjorken} labeled by a $+$ vanish and the gradient expansion converges provided that 
\begin{equation}
\textrm{Kn} < \textrm{Kn}^* \equiv \frac{3\sqrt{2}}{a\sqrt{b}}.   
\end{equation}
This existence of a critical Knudsen number is  qualitative agreement with the findings of Ref.\,\cite{Denicol:2019lio}. The quantitative difference in the value of $\textrm{Kn}^*$ reported here and the one originally quoted in Ref.\,\cite{Denicol:2019lio} is explained by the fact that our gradient expansion is different from the one considered there. 

We conclude this Appendix by arguing in two different ways that the large-order geometric growth displayed by $\Pi_\star^{(n)}$ can be understood as a fine-tuned phenomenon. 

The first way starts from the realisation that, had one chosen to write the closed-form solution \eqref{DN:Bjorken_recursion-relation_solution} in terms of an expansion in the longitudinal derivative $\partial_\tau$ instead of $\tau \partial_\tau$, one would had gotten an expression of the form
\begin{equation}
\Pi_\star^{(n)} = \sum_{k=0}^{n-1} c_{n,k} \tau^{k-1} \partial_\tau^k T(\tau),
\end{equation}
where, in particular, 
\begin{equation}
c_{n,n-1} = \left(-\frac{1}{4}a b \textrm{Kn}\right)^{n-1} \frac{2a}{3\sigma_T}.\label{cn} 
\end{equation}
For a $T(\tau)$ given by a linear combination of power-laws $\tau^{-\lambda}$, the contribution to the gradient expansion of the form 
\begin{equation}
\Pi_\star \supset \sum_{n=1}^\infty c_{n,n-1} \tau^{n-2}\partial_\tau^{n-1} T(\tau),
\end{equation}
is a factorially divergent asymptotic series when evaluated at a given time since according to Eq.\,\eqref{cn} the coefficients $c_{n,n-1}$ grow geometrically and 
\begin{equation}
\partial_\tau^k \tau^{-\lambda} = (-1)^k \tau^{-\lambda-k} \frac{\Gamma(\lambda+k)}{\Gamma(\lambda)}.
\end{equation}
The fact that the whole gradient expansion only grows geometrically when adding the remaining contributions is indicative of a very delicate cancellation between different factorially growing subsectors.

The second way builds upon an observation already performed in the main text: $\Pi_\star^{(1)}$ is a linear superposition of eigenfunctions of the differential operator $\mathcal{M}$. Let us elaborate on this. According to the definition \eqref{M_def} and the recursion relation \eqref{DN_rec_rel_tau_2}, 
\begin{equation}
\mathcal{M} = {-}\frac{a b \textrm{Kn}}{3}{-}\frac{a b \textrm{Kn}}{4} \tau\partial_\tau,
\end{equation}
in such that a way that its eigenfunctions $\psi_i$, $\mathcal{M}\psi_i = \xi_i \psi_i$, are given by 
\begin{equation}
\psi_i = \tau^{-\frac{4}{3}-\frac{4}{a b \textrm{Kn}}\xi_i}.
\end{equation}
For a $\Pi_\star^{(1)}$ corresponding to a finite linear superposition of eigenfunctions,
\begin{equation}\label{DN_Pi1_spectral}
\Pi_\star^{(1)} = \sum_i c_i \psi_i,
\end{equation}
the gradient expansions grows geometrically with rates given by $\xi_i$,
\begin{equation}
\Pi_\star^{(n+1)} = \mathcal{M}^n\Pi_\star^{(1)} = \sum_i \xi_i^n c_i \psi_i.
\end{equation}
Eqs.\,\eqref{DN_Pi1_spectral} and \eqref{DN_rec_rel_tau_1} imply that
\begin{equation}\label{DN_T_spectral}
T = \frac{3}{2 a \rho_0 \tau_0 \textrm{Kn}}\sum_i c_i \tau^{-\frac{1}{3}-\frac{4}{a b \textrm{Kn}}\xi_i}.
\end{equation}
Inserting the expression above in the conservation equation fixes $\Pi_\star$, 
\begin{equation}\label{DN_Pi*_spectral}
\Pi_\star = -\frac{9}{a^2 b \textrm{Kn}^2}\sum_i c_i \xi_i \tau^{-\frac{4}{3}-\frac{4}{a b \textrm{Kn}}\xi_i}.
\end{equation}
Finally, for the whole procedure to be self-consistent, the $\Pi_\star$ given by Eq.\,\eqref{DN_Pi*_spectral} has to solve the relaxation equation. This can only be achieved provided that the linear superposition \eqref{DN_Pi*_spectral} is restricted to two contributions with eigenvalues 
\begin{equation}
\xi_{1,2} = \frac{1}{2}\pm\frac{1}{6}\sqrt{9+4a^2b\textrm{Kn}^2}.
\end{equation}
The eigenvalues $\xi_{1,2}$ agree with the growth rates quoted in Eq.\,\eqref{DN_gamma_pm} and, when introduced in Eq.\,\eqref{DN_T_spectral} (resp. Eq.\,\eqref{DN_Pi*_spectral}), match the exponents appearing in Eq.\,\eqref{DN_alpha_pm} (resp. Eq.\,\eqref{DN_grad_exp_Bjorken_1}). 

In a generic model with a recursion relation of the form \eqref{M_def} $\mathcal{M}$ depends nontrivially on the hydrodynamic fields. In this situation, while a $\Pi_\star^{(1)}$ given by a finite superposition of eigenfunctions would result in a geometrically growing gradient expansion, demanding that the on-shell hydrodynamic fields conspire to produce precisely a $\Pi_\star^{(1)}$ of this form is a tall order. It is natural to expect that this is not the case for a generic flow, providing another sense in which the geometric growth of the gradient expansion for Bjorken flow in the DN model is a fine-tuned phenomenon.

\end{document}